\documentclass{article}
\usepackage[english]{babel}
\usepackage[latin1]{inputenc}
\usepackage{amsmath}
\usepackage{hhline}
\usepackage{amssymb}
\usepackage{latexsym}
\usepackage{xspace}
\usepackage{fancybox}
\usepackage{multicol}
\usepackage{pstricks}
\usepackage{pst-node}
\newcommand{\eop}{\hbox{\hskip6pt\vrule height 6pt width 6pt}}
\newcommand{\ignore}[1]{}
\newcommand{\boxtheorem}{\hfill $\Box$}
\newcommand{\nit}[1]{{\it #1}}

\newcommand{\rpr}[2]{{\it Repairs\/}_{#1}({#2})}
\newcommand{\core}[2]{{\it Core\/}_{#1}({#2})}

\newtheorem{definition}{Definition}
\newtheorem{theorem}{Theorem}

\newtheorem{proposition}{Proposition}
\newtheorem{lemma}{Lemma}

\newtheorem{example}{Example}

\newenvironment{proof}{\noindent{\em Proof.}}{\eop}

\title{\bf On the Computational Complexity of Consistent Query Answers}

\author{{\bf Jan Chomicki}\thanks{Contact author. Address: Dept. CSE,
201 Bell Hall, Univ. at Buffalo, Buffalo, NY 14260-2000. Fax: (716) 645-3464.
Phone: (716) 645-3180, ext.103.}\\
Dept. CSE\\
University at Buffalo\\
chomicki@cse.buffalo.edu\\
\and {\bf Jerzy Marcinkowski}\\
Instytut Informatyki\\
Wroclaw University, Poland\\
jma@ii.uni.wroc.pl\\
}
\date{}

\begin{document}
\maketitle
\begin{center}
\end{center}

\ignore{
\begin{abstract}

We consider here the problem of obtaining reliable, consistent
information from inconsistent databases -- databases that do not have
to satisfy given integrity constraints. We use the notion of
consistent query answer -- a query answer which is true in every
(minimal) repair of the database.  We provide a complete
classification of the computational complexity of consistent answers
to first-order queries w.r.t.  functional dependencies and denial
constraints.  We show how the complexity depends on the {\em type} of
the constraints considered, their {\em number}, and the {\em size} of
the query.  We obtain several new PTIME cases, using new algorithms.
\end{abstract}
}
\section{Introduction}
\label{sec:intro}

It is nowadays common 
to build databases integrating information from multiple, autonomous,
distributed 
data sources. The problem of {\em data integration} is nevertheless
very complex \cite{Hull97,Mot99}. 
In this paper, we consider a specific issue arising in data integration;
how to obtain reliable, consistent information from {\em
inconsistent databases} -- databases that do not have to satisfy given
integrity constraints. Such databases occur in a natural way in data
integration, since there is typically no global monitor that could guarantee
that the integrated database satisfies the constraints. The data
sources are independent and even if they separately satisfy the
constraints, the integrated database may fail to do so. For example,
different data sources may contain different, locally unique addresses
for the same person, leading to the violation of the global uniqueness
constraint for people's addresses. 
Inconsistent databases occur also in other contexts. For instance,
integrity constraints may fail to be enforced for efficiency reasons,
or because the inconsistencies  are temporary.
Or, there may be not enough information to resolve inconsistencies,
while the database may have to continue being used for real-time
decision support. 

To formalize the notion of consistent information obtained from a (possibly
inconsistent) database in response to a user query, we proposed in \cite{ArBeCh99} 
the notion of a {\em consistent query answer}. A consistent answer is,
intuitively, true regardless of the way the database is fixed
to remove constraint violations.
Thus answer consistency serves as an indication of its reliability.
The different ways of fixing an inconsistent database are formalized using
the notion of {\em repair}: another database that is consistent and minimally differs from
the original database.
\begin{example}
Consider the following instance of a relation {\it Person}
\begin{center}
\begin{tabular}{|lll|}
\hline
{\it Name} & {\it City} & {\it Street}\\\hline
{\rm Brown} & {\rm Amherst} & {\rm 115 Klein}\\
{\rm Brown} & {\rm Amherst} & {\rm 120 Maple}\\
{\rm Green} & {\rm Clarence} & {\rm 4000 Transit}\\\hline
\end{tabular}
\end{center}

\noindent
and the functional dependency ${\it Name}\rightarrow {\it City}\;{\it Street}$.
Clearly, the above instance does not satisfy the dependency.
There are two repairs: one is obtained by removing the first tuple,
the other by removing the second.
The consistent answer to the query $\;{\it Person\/}(n,c,s)\;$
is just the tuple \mbox{\rm (Green,Clarence,4000 Transit)}.
On the other hand, the query $\;\exists s[{\it Person\/}(n,c,s)]\;$
has two consistent answers: \mbox{\rm (Brown,Amherst)} and 
\mbox{\rm (Green,Clarence)}.
Similarly, the query 
\[{\it Person\/}({\rm Brown},{\rm Amherst},{\rm 115\; Klein})
\vee {\it Person\/}({\rm Brown},{\rm Amherst},{\rm 120\; Maple})\]
has {\em true} as the consistent answer.
Notice that for the last two queries the approach based on removing
all inconsistent tuples and evaluating the original query using
the remaining tuples gives different, less informative results.
\end{example}

In \cite{ArBeCh99}, in addition to a formal definition of a consistent
query answer, a computational mechanism for obtaining such answers was
presented in the context of first-order queries. In \cite{ArBeCh01},
the same problem was studied for scalar aggregation queries. 
In \cite{ArBeCh99} some  cases  were identified where consistent query answers 
are tractable (in PTIME).
In the present paper, we provide a {\em complete classification} of the computational
complexity of computing consistent query answers to
first-order queries. We consider {\em functional dependencies} (FDs)
and their generalization: {\em denial constraints}. 
Denial constraints allow an arbitrary number of literals per constraint
and arbitrary built-in predicates. They also relax the typedness
restriction of FDs.
Denial constraints are particularly useful for databases with
interpreted data, e.g., numbers.
Their implication problem was studied in \cite{BaChWo99}.
\begin{example}
The constraint that {\em no employee can have a salary greater than that of
her manager} is the denial constraint
\[\forall n,s,m,s',m'. \neg [{\it Emp\/}(n,s,m)\wedge {\it Emp\/}(m,s',m')\wedge
s>s'].\]
\end{example}

The results of \cite{ArBeCh99} imply that for binary denial constraints
consistent answers can be computed in PTIME for queries that are conjunctions of literals.
In the present paper we strengthen that result to arbitrary quantifier-free
queries and arbitrary denial constraints. We also identify a class of restricted
existentially quantified queries (consisting of single literals), for which
consistent query answers can be computed in PTIME.
In general, we show how the complexity depends on the {\em type} of the constraints
considered, their {\em number}, and the {\em size} of the query.
Related work is discussed in depth in \cite{ArBeCh99,ArBeCh01,ABCHRS03}.
Other papers that adopt our notion of consistent query answer include
\cite{ArBeCh00,GrZu00,GrGrZu01}.

\section{Basic Notions}\label{sec:basic}

In this paper we assume that we have a fixed database schema
containing only one relation schema $R$ with the set of attributes
$U$. We will denote elements of $U$ by $A,B,\ldots$, subsets
of $U$ by $X,Y,\ldots$, and the union of $X$ and $Y$ by $XY$.
We also have two fixed, disjoint
infinite database domains: $D$ (uninterpreted
constants) and $N$ (numbers).
We assume
that elements of the domains with different names are different.  The
database instances can be seen as first order structures that
share the domains $D$ and $N$. 
Every attribute in $U$ is typed, thus all the instances of $R$
can contain only elements either of $D$ or of $N$ in a single attribute.
Since each instance is finite, it
has a finite active domain which is a subset of $D\cup N$.  As usual, we allow
the standard built-in predicates over $N$ ($=,\not=,<,>,\leq,\geq$)
that have infinite, fixed  extensions.
\ignore{
The assumption of a single relation schema is inessential for the upper
bounds and the presence of the number domain is not necessary for the lower bounds to hold.
}

Integrity constraints are typed, closed first-order formulas over the
vocabulary consisting of $R$ and the built-in predicates over $N$.

\begin{definition}
Given a database instance $r$ of $R$ and a set of integrity constraints $F$,
we say that $r$ is {\em consistent} if $r\vDash F$ in the standard model-theoretic
sense; {\em inconsistent} otherwise.
\end{definition}

We consider the following classes of integrity constraints:
\begin{itemize}
\item {\em denial constraints}: formulas of the form
$\forall\bar{x}_1,\ldots\bar{x}_k. \neg[R(\bar{x}_1)\wedge\cdots\wedge R(\bar{x}_m)
\wedge \phi(\bar{x}_1,\ldots,\bar{x}_m)]$
where $\bar{x}_1,\ldots,\bar{x}_m$ are tuples of variables and constants, and $\phi$
is a conjunction of atomic formulas referring to built-in predicates;
\item {\em functional dependencies} (FDs) $X\rightarrow Y$ over the set $U$
({\em key} FDs if $X$ is a key of $R$).
\end{itemize}

Clearly, functional dependencies are a special case of denial constraints.

\begin{definition} \label{def:order}
For the instances $r,r',r'' ~$, $r'\leq_r r''$ if $r-r'\subseteq r-r''$.
\ignore{
$\Delta(r,r')\subseteq\Delta(r,r'')$, i.e., if the distance between
$r$ and $r'$ is less than or equal to the distance between $r$ and $r''$.}
\boxtheorem
\end{definition}

\begin{definition} 
Given a set of integrity constraints $F$ and database instances $r$ and $r'$, we say that
$r'$ is a {\em repair} of $r$ w.r.t. $F$ if $r'\vDash F$ and $r'$ is
$\leq_r$-{\it minimal} in the class of database instances that satisfy
$F$. \boxtheorem
\end{definition}

We denote by $\rpr{F}{r}$ the set of repairs of $r$ w.r.t. $F$.
For any set of denial constraints, all the repairs are  obtained by deleting tuples from the table.
\ignore{
Also, in this case the union of all repairs of any instance 
$r$ is equal to $r$. These properties are not necessarily
shared by other classes of ICs.
}
\ignore{
\begin{definition}
The {\em core} of $r$ is defined as~
\[\core{F}{r}=\bigcap_{r'\in \rpr{F}{r}} r'.\] 
\hfill $\Box$
\end{definition}

The {\em core} is a new database instance. If $r$ consists of a single
relation, then the core is the intersection of all
the repairs of $r$. The core of $r$ itself is not necessarily a repair of $r$.
}

\ignore{
Query answers for first order queries are defined in the
standard way.
}
\begin{definition}\cite{ArBeCh99}\label{def:cqa}
Given a set of integrity constraints $F$ and a database instance $r$,
we say that a (ground) tuple $\bar{t}$ is a {\em consistent answer} to a query
$Q(\bar{x})$ w.r.t. $F$
in $r$, and we write
~$r \models_F Q(\bar{t})$ 
if for every $r'\in\rpr{F}{r}$,  $r'\vDash\ Q(\bar{t})$.
If $Q$ is a sentence, then $\nit{true}$ ($\nit{false}$) is a
{\em consistent answer} to $Q$ w.r.t. $F$ in $r$, and we write $r \models_F Q$
~($r \models_F \neg Q$), if for every $r'\in\rpr{F}{r}$,
$r' \vDash Q$ ~($r' \nvDash Q$).
\boxtheorem
\end{definition}

\section{Data Complexity of Consistent Query Answers}
Assume a class of databases ${\cal D}$, a class of queries ${\cal L}$ and a
class of integrity constraints ${\cal IC}$ are given.
We study here the {\em data complexity} \cite{ChHa80,Var82}
of consistent query answers, i.e., the 
complexity of (deciding the membership of) the sets
$D_{F,\phi}=\{r:r\in {\cal D} \wedge r\models_F\phi\}$ for a fixed sentence
$\phi\in{\cal L}$ and a fixed finite set $F\in{\cal IC}$ of integrity constraints.

\begin{proposition}\cite{ABCHRS03}\label{c:foquery}
For any set of denial constraints $F$ and sentence $\phi$,
$D_{F,\phi}$ is in co-NP.
\end{proposition}

It is easy to see that even under a single key FD, there may be
exponentially many repairs and thus the approach to computing consistent
query answers by generating and examining all repairs is not feasible.

\begin{example}\label{ex:expo}
Consider the functional
dependency $A\rightarrow B$ and
the following  family of relation instances
$r_n$, $n>0$, each of which has $2n$ tuples (represented as columns)
and $2^n$ repairs:
\vspace{-0.6cm}
\begin{center}
\[\begin{array}{c|ccccccc}
r_n & &&&&&&\\ \hline
A & a_1& a_1& a_2& a_2 & ~~\cdots &a_n &a_n\\ \hline
B & b_0& b_1& b_0& b_1 & ~~\cdots &b_0 &b_1 \\ \hline
\end{array}\]
\end{center}
\ignore{
\begin{center}
\[\begin{array}{|ll|}
\hline 
A&B\\\hline
1&0\\
1&1\\
2&0\\
2&1\\
\multicolumn{2}{c}{\cdots}\\
n&0\\
n&1\\\hline
\end{array}\]
\end{center}
}
\end{example}


Given a set of denial constraints $F$ and an instance $r$, all the repairs of $r$
with respect to $F$ can be succinctly represented as the {\em conflict hypergraph}.
This is a generalization of the {\em conflict graph} defined in \cite{ArBeCh01}
for FDs only.
\begin{definition}\label{d:hypergraph}
The {\em conflict hypergraph} ${\cal G}_{F,r}$ is a hypergraph
whose set of vertices is the set of tuples in $r$ and
whose set of edges consists of all the sets $\{\bar{t}_1,\bar{t}_2,\ldots \bar{t}_l\}$
such that $\bar{t}_1,\bar{t}_2,\ldots \bar{t}_l\in {r}$, and there is a constraint
\[\forall \bar{x}_1,\bar{x}_2,\ldots \bar{x}_l 
\neg [R(\bar{x}_1) \wedge R(\bar{x}_2) \wedge \ldots \wedge R(\bar{x}_l)\wedge 
\phi(\bar{x}_1,\bar{x}_2,\ldots \bar{x}_l)]\]
in $F$ such that $\bar{t}_1,\bar{t}_2,\ldots \bar{t}_l$ 
violate together  this constraint, which means that there exists 
a substitution $\rho$ such that 
$\rho(\bar{x}_1)=\bar{t}_1, \rho(\bar{x}_2)=\bar{t}_2, \ldots \rho(\bar{x}_l)=\bar{t}_l$ and that 
$\phi(\bar{t}_1,\bar{t}_2,\ldots \bar{t}_l) $ is true.
\end{definition}

By an {\em independent set} in a hypergraph we mean a subset of its set of vertices
which does not contain any edge.  

\begin{proposition}
Each repair of $r$ w.r.t. $F$  corresponds to
a maximal independent set in ${\cal G}_{F,r}$.
\end{proposition}

\subsection{Positive results}

\begin{theorem}
For every set $F$  of denial constraints and quantifier-free sentence $\Phi$, $D_{F,\Phi}$ is in PTIME.
\end{theorem}

\begin{proof}
We assume the sentence is in CNF, i.e., of the form
$\Phi = \Phi_1\wedge \Phi_2 \wedge \ldots \Phi_l $, where  each $\Phi_i $ is a disjunction of
ground literals. $\Phi$ is true in every repair of $r$ if and only if each of the clauses $\Phi_i $ is
true in every repair. So it is enough to provide a polynomial algorithm which will check if for  
a given  ground clause {\em true} is a consistent answer. 

It is easier to think that we are checking if for a ground clause {\em true} is {\bf not } a consistent answer. 
This means that we are checking, whether there exists a repair $r'$ in which $\neg \Phi_i$ is true
for some $i$. 
But $\neg \Phi_i $ is of the form 
$ R(\bar{t}_1)\wedge R(\bar{t}_2)  \wedge \ldots \wedge  R(\bar{t}_m)\wedge \neg R(\bar{t}_{m+1})\ldots \neg 
R(\bar{t}_n)$, 
where  $\bar{t}_j$ are  tuples of constants.
\ignore{
First we show a nondeterministic algorithm which
tries to build the repair {r'}: it keeps in {r'} all the tuples $\bar{t_1},\ldots \bar{t_m}$. Then, for each
$m+1\leq j \leq n $ it selects one edge of the conflict hypergraph ${\cal G}_{F,r}$ containing the vertex $\bar{t_j}$
(if, for some $\bar{t_j}$, there are no such edges then the algorithm fails), 
and it adds to $r'$ all the vertices of this edge, except of $\bar{t_j}$. If, at this stage,
there exists an edge
of ${\cal G}_{F,r}$ whose all vertices are in $r'$ then the algorithm fails. 
}
\ignore{
The condition $(*)$ can be checked
by a  nondeterministic algorithm  that needs  $n-m$ nondeterministic steps, a number which is independent of
the  size of the database, and in each of its nondeterministic steps selects one possibility from a set
whose size is polynomial in the size of the database. 
So there is an equivalent PTIME deterministic algorithm.
}
Thus it is enough to check two conditions:
\begin{enumerate}
\item whether for every $j$, $m+1\leq j \leq n$, $\bar{t}_j\not\in r$ or
there exists an edge $E_j\in {\cal G}_{F,r}$ such that $\bar{t}_j\in E_j$, and
\item there is no edge $E\in {\cal G}_{F,r}$ such that
$E\subseteq {r'}$
where
\[{r'}= \{\bar{t}_1,\ldots,\bar{t}_m\}\cup \bigcup_{m+1\leq j\leq n,\bar{t}_j\in r} (E_j-\{\bar{t}_j\}).\]
\end{enumerate}
If the conditions are satisfied, then 
a repair in which $\neg \Phi_i$ is  true can be built by adding to $r'$ new tuples 
from $r$ until the set is maximal independent. 
\ignore{
Thus it is enough to check whether 
$\{\bar{t}_1,\ldots,\bar{t}_m\}\subseteq {r}$ and whether 
there exists a family $\{E_j\}_{m+1\leq j \leq n}$  of edges of $ {\cal G}_{F,r}$ such that
$\bar{t}_j\in E_j$,
and that there is no edge $E\in {\cal G}_{F,r}$ such that
$E\subseteq {r'}$
where
\[{r'}= \{\bar{t}_1,\ldots,\bar{t}_m\}\cup \bigcup_{m+1\leq j\leq n}
(E_j-\{\bar{t}_j\}).\]
If the condition $(*)$ is satisfied, then 
a repair in which $\Phi_i$ is not true can be built by adding to ${r'}$ new tuples 
from {r} until the set is maximal independent. 
}
The conditions can be checked
by a  nondeterministic algorithm  that needs  $n-m$ nondeterministic steps, a number which is
independent of
the  size of the database, and in each of its nondeterministic steps selects one possibility from a
set
whose size is polynomial in the size of the database. 
So there is an equivalent PTIME deterministic algorithm.
\end{proof} 
 
Note that the above result holds also for constraints and queries involving more than one relation.
The notion of conflict hypergraph needs to be appropriately generalized in this case.


\ignore{

In the case when the set $F$ of integrity constraints consists of only one FD the
conflict graph has a very simple form.  It is a disjoint union of full multipartite graphs. 
If this single dependency is a key dependency then the conflict graph is a union of disjoint cliques. 
Because of this very simple structure we hoped that it would be possible, in such a situation, 
to compute in polynomial time the consistent answers not only to
ground queries, but also to all  conjunctive queries. As we are going to see, this is not the case.
The strongest positive result is the following:
}
\begin{theorem}
Let $F$ consist of a single FD. 
Then for each sentence $Q$   of the form 
$\exists \bar{t} [R(\bar{t})\wedge \phi(\bar{t})]$
(where  $\phi$ is quantifier-free and only built-in predicates occur there),
there exists a sentence $Q'$ such that for every database instance {r},
$r\models_F Q$ 
iff $r\models Q'$.
Consequently, $D_{F,Q}$ is 
in PTIME.
\end{theorem}

\ignore{
Notice that we can force some of the $x_i$ in the query 
above to be fixed constants by using in $\phi$ equations of the
form $x_i=c$. 
}
\begin{proof}
The FD is $A_1\ldots A_l\rightarrow A_{l+1},\ldots A_{l+m}$, where 
$l+m$ is not greater than the arity $k$ of $R$.
Let $\bar{x}$  be a vector of distinct variables of length $l$,
$\bar{y}$ and $\bar{y}_1$ vectors of distinct variables of length $m$,
and $\bar{z}$, $\bar{z}_1$ and $\bar{z}_2$ vectors of distinct variables of length $k-(l+m)$.
Then, the query $Q'$ is as follows:
\[\exists \bar{x},\bar{y},\bar{z}\forall \bar{y}_1,\bar{z}_1\exists \bar{z}_2
[R(\bar{x},\bar{y},\bar{z})\wedge\phi(\bar{x},\bar{y},\bar{z})\wedge
[R(\bar{x},\bar{y}_1,\bar{z}_1)\Rightarrow [R(\bar{x},\bar{y}_1,\bar{z}_2)\wedge
\phi(\bar{x},\bar{y}_1,\bar{z}_2)]]].\]
\end{proof}

We show now that the above results are the strongest possible, since
relaxing any of the restrictions leads to co-NP-completeness.
This is the case even though we limit ourselves to {\em key} FDs.

\subsection{One key dependency, two query literals}

\ignore{
\begin{lemma}\label{zksiazeczki}
3SAT is NP-complete even if restricted only to instances, in which every clause
either contains only positive literals or only negative literals.
\end{lemma}
\begin{proof} For each variable $p$ 
substitute every occurrence of $\neg p$ by a new variable $p'$. Then add clauses
$p\vee p'$ and $\neg p\vee \neg p'$.\end{proof}
}

\begin{theorem}\label{bliskosiebie}
There exist a key FD $f$ and a query $Q\equiv\exists x,y,z [R(x,y,c)\wedge R(z,y,c')]$,
for which $D_{\{f\},Q}$ is co-NP-data-complete.
\end{theorem}

\begin{proof}
Reduction from MONOTONE 3-SAT.
The FD is $A\rightarrow BC$.
Let $\Phi=\phi_1\wedge\ldots \phi_m \wedge \psi_{m+1} \ldots \wedge\psi_l $ be a conjunction of clauses, such that
all occurrences of variables in $\phi_i$ are positive and all occurrences of variables in $\psi_i$ are negative.
We build a database with  the facts $R(i,p,c)$ if the variable $p$ occurs in the clause $\psi_i$
and  $R(i,p,c')$ if the variable $p$ occurs in the clause $\phi_i$. 
Now, there is an assignment which satisfies $\Phi$  
if and only if  there exists a repair 
of the database in which $Q$ is false. 
To show  the $\Rightarrow$ implication, select for each clause $\phi_i$ one variable $p_i$
which occurs in this clause and whose value is 1 and for each clause $\psi_i$  one variable $p_i$
which occurs in $\psi_i$ and whose value is 0. The set of facts 
$\{R(i,p_i,c):i\leq m\}\cup  \{R(i,p_i,c'):m+1\leq i\leq l\}$ is a repair in which the query $Q$ is false. 
The $\Leftarrow$ implication is even simpler.
\end{proof}

\subsection{Two key dependencies, one query literal}

By a {\em bipartite edge-colored graph} we mean  a tuple ${\cal G}=\langle V,E,B,G\rangle $ such that
 $\langle V,E\rangle$ is an undirected bipartite graph and  $E=B\cup G $ for some given 
 disjoint  sets $B,G$ (so we think that each of the edges of $\cal G$ has one
of the two colors). 

\begin{definition}
Let  ${\cal G}=\langle V,E,B,G\rangle $ be a  bipartite edge-colored graph, 
and let $F\subset E$.  
We say that $F$ is maximal $\cal V$-free if:

\begin{enumerate}
     \item $F$ is a maximal (w.r.t. inclusion) subset of $E$ with the property
      that neither $F(x,y)\wedge F(x,z)$ nor $F(x,y)\wedge F(z,y)$ holds
      for any $x,y,z$.

     \item $F\cap B =\emptyset$.
\end{enumerate} 

We say that $\cal G$ 
has the max-$\cal V$-free property
if there exists $F$ which is maximal $\cal V$-free.
\end{definition}

\begin{lemma}\label{2key}
Max-$\cal V$-free is an NP-complete property of bipartite edge-colored graphs.
\end{lemma}
\begin{proof}
Reduction from 3-COLORABILITY. Let ${\cal H}=\langle U,D\rangle $ be some 
undirected graph. 
This is how we define the bipartite edge-colored graph  ${\cal G}_{\cal H}$:

\begin{enumerate}
\item
$V=\{v_\varepsilon, v'_\varepsilon :v\in U,\varepsilon\in\{m,n,r,g,b\}\}$,
which means that there are 10 nodes in the graph $\cal G$ for each node of $\cal H$;

\item
$G(v_m,v'_r),G(v_m,v'_b),G(v_n,v'_b),G(v_n,v'_g)$ and
$G(v_r,v'_m)$,$G(v_b,v'_m)$,$G(v_b,v'_n)$,$G(v_g,v'_n)$ hold for each $v\in U$;

\item
$B(v_\epsilon,v'_\varepsilon) $ holds for each $v\in U$ and each pair 
$\epsilon,\varepsilon\in\{r,g,b\}$ such that $\epsilon \neq \varepsilon $;

\item
$B(v_\varepsilon,u'_\varepsilon) $ holds for each $\varepsilon\in \{r,g,b\}$
and each  pair $u,v\in U$  such that $ D(u,v) $.

\end{enumerate}

Suppose that $\cal H$ is 3-colorable. We fix a coloring of $\cal H$ and construct the set $F$.
 For each $v\in U$: if the color of $v$ is Red, then  the edges 
$G(v_m,v'_b),G(v_n,v'_g)$ and $G(v_b,v'_m),G(v_g,v'_n)$ are in $F$.
If color of $v$ is Green, then  
 the edges 
$G(v_m,v'_r),G(v_n,v'_b)$ and $G(v_r,v'_m), G(v_b,v'_n)$
are in $F$, and if  the color of $v$ is Blue, then  
 the edges 
$G(v_m,v'_r),G(v_n,v'_g)$ and $G(v_r,v'_m),G(v_g,v'_n)$ are in $F$. It is easy to see that 
the set $F$ constructed in this way is maximal $\cal V$-free.

For the other direction, suppose that a maximal $\cal V$-free set $F$ exists in ${\cal G}_{\cal H}$.
Then, for each $v\in U$ there is at least one  node among $v_r,v_g, v_b$ which does not belong 
to any $G$-edge in $F$. Let $v_\epsilon$ be this node.
Also, there is at least one such node (say, $v'_\varepsilon$) among $v'_r,v'_g, v'_b$. Now, it follows easily
from the construction of ${\cal G}_{\cal H}$ that if $F$ is  
maximal $\cal V$-free then $\epsilon = \varepsilon$. Let this $\epsilon $ be color of $v$ in $\cal G$. 
It is easy to check that the coloring  defined in this way is a legal 3-coloring of $\cal G$.
\end{proof}

\begin{theorem}
There is a set $F$ of 2 key dependencies and a query $Q\equiv\exists x,y [R(x,y,b)]$, 
for which $D_{F,Q}$ is co-NP-data-complete.
\end{theorem}

\begin{proof}
The 2 dependencies are $A\rightarrow BC$ and $B\rightarrow AC$. For a given 
 bipartite edge-colored graph  ${\cal G}=\langle V,E,B,G\rangle $ we build a database with 
 the tuples $(x,y,g)$ if $G(x,y) $ holds in ${\cal G}$ and $(x,y,b)$ if $B(x,y) $ holds in ${\cal G}$.
 Now the theorem follows from Lemma \ref{2key} since
 a repair in which  the query $\exists x,y \; R(x,y,b)$ is not true exists if and only if 
 ${\cal G}$ has the max-$\cal V$-free property. \end{proof}

\subsection{One denial constraint}

By an {\em edge-colored graph} we mean  a tuple ${\cal G}=\langle V,E,P,G,B\rangle $ such that
 $\langle V,E\rangle$ is a (directed) graph and  $E=P\cup G \cup B$ for some given 
pairwise disjoint  sets $P,G,B$ (which we interpret as colors). 
We say that the edge colored graph $\cal G$ has the $\cal Y$ property if there are 
$x,y,z,t\in E$ such that 
$E(x,y),E(y,z),E(y,t)$ hold and the edges $E(y,z)$ and $E(y,t)$ are of different colors.

\begin{definition}\label{tenF}
We say that the edge-colored graph $\langle V,E,P,G,B\rangle$ has the max-$\cal Y$-free property
if there exists a subset $F$ of $E$ such that $F\cap P =\emptyset$ and :
\begin{enumerate}
\item $\langle V,F,P\cap F,G\cap F,B\cap F \rangle $ does not have the $\cal Y$-property;

\item $F$ is a maximal (w.r.t. inclusion) subset of $E$ satisfying the first condition;

\end{enumerate} 
\end{definition}

\begin{lemma}\label{1denial}
Max-$\cal Y$-free is an NP-complete property of edge-colored graphs.
\end{lemma}
\begin{proof}
By a reduction of 3SAT. Let $\Phi= \phi_1\wedge \phi_2\wedge \ldots \wedge \phi_l$ be 
conjunction of clauses. Let $p_1,p_2,\ldots p_n$ be all the variables in $\Phi$.
This is how we define the  {\em edge-colored graph} ${\cal G}_{\Phi}$: 

\begin{enumerate}
\item 
$V=\{a_i,b_i,c_i,d_i:1\leq i\leq n\}\cup \{e_i,f_i,g_i:1\leq i\leq l\}$,
which means that there are 3 nodes in the new graph for each clause in $\Phi$
and 4 nodes for each variable. 

\item $P(a_i,b_i)$ and $P(e_j,f_j)$ hold for each suitable $i,j$;

\item $G(b_i,d_i)$ and $G(e_j,g_j)$ hold for each suitable $i,j$;

\item $B(b_i,c_i)$  holds for each suitable $i$;

\item $G(d_i,e_j)$ holds if $p_i$ occurs positively in $\phi_j$;

\item $B(d_i,e_j)$ holds if $p_i$ occurs negatively in $\phi_j$;

\item  $E=B\cup G\cup P$.
\end{enumerate}

Now suppose that $\Phi $ is satisfiable, and that $\mu$ is the 
satisfying assignment. We define the set $F\subset E$ as follows.
We keep in $F$ all the $G$-colored edges from item 3 above. If $\mu(p_i)=1$
then we keep in $F$ all the $G$ edges leaving $d_i$ (item 5). Otherwise we  
keep in $F$ all the $B$ edges leaving $d_i$ (item 6). Obviously, $F\cap P=\emptyset$.
It is also easy to see that $F$ does not have the $\cal Y$-property and that it is maximal.

In the opposite direction, notice that if an $F$, as in Definition \ref{tenF} does exist, then
it  must contain
all the $G$-edges from item 2 above - otherwise a $P$ edge could be added without 
leading to the $\cal Y$-property. But this means that, for each $i$,
$F$ can either contain some (or all) of the $B$-edges leaving $d_i$ or some (or all) of the $G$-edges.
In this sense $F$ defines a valuation of variables. Also, if $F$ is maximal, it must contain,
for each $j$, at least one edge leading to $e_j$. But this  means that the defined valuation 
satisfies $\Phi$.
\end{proof}


\begin{theorem}
There exist a denial constraint $f$ and a query of the form $Q\equiv\exists x,y [R(x,y,p)]$, 
for which $D_{\{f\},Q}$ is co-NP-data-complete.
\end{theorem}

\begin{proof}
The denial constraint $f$ is:
\begin{center}
 $\forall x,y,z,s,s',s''\; \neg[ R(x,y,s) \wedge R(y,z.s') \wedge R(y,w,s'')\wedge s'\neq s'']$
\end{center}
 For a given  edge-colored graph  ${\cal G}=\langle V,E,P,G,B\rangle $ we build a database with 
 the tuples $R(x,y,g)$ if $G(x,y) $ holds in ${\cal G}$, with  $R(x,y,p)$ if $P(x,y) $ holds in ${\cal G}$
 and with $R(x,y,b)$ if $B(x,y) $ holds in ${\cal G}$.
 Now the theorem follows from Lemma \ref{1denial} since
 a repair in which  the query $Q$ is not true exists iff
 ${\cal G}$ has the max-$\cal Y$-free property. \end{proof}

\vspace{2mm}\noindent
\section{Acknowledgment}
Work supported in part by NSF grant IIS-0119186.


\begin{thebibliography}{10}

\bibitem{ArBeCh99}
M.~Arenas, L.~Bertossi, and J.~Chomicki.
\newblock {Consistent Query Answers in Inconsistent Databases}.
\newblock In {\em ACM Symposium on Principles of Database Systems}, pages
  68--79, 1999.

\bibitem{ArBeCh00}
M.~Arenas, L.~Bertossi, and J.~Chomicki.
\newblock {Specifying and Querying Database Repairs Using Logic Programs with
  Exceptions}.
\newblock In {\em International Conference on Flexible Query Answering
  Systems}, pages 27--41. Springer-Verlag, 2000.

\bibitem{ArBeCh01}
M.~Arenas, L.~Bertossi, and J.~Chomicki.
\newblock {Scalar Aggregation in FD-Inconsistent Databases}.
\newblock In {\em International Conference on Database Theory}, pages 39--53.
  Springer-Verlag, LNCS 1973, 2001.

\bibitem{ABCHRS03}
M.~Arenas, L.~Bertossi, J.~Chomicki, X.~He, V.~Raghavan, and J.~Spinrad.
\newblock {Scalar Aggregation in Inconsistent Databases}.
\newblock {\em Theoretical Computer Science}, 2003.
\newblock Special issue: selected papers from ICDT 2001, to appear.

\bibitem{BaChWo99}
M.~Baudinet, J.~Chomicki, and P.~Wolper.
\newblock {Constraint-Generating Dependencies}.
\newblock {\em Journal of Computer and System Sciences}, 59:94--115, 1999.
\newblock Preliminary version in ICDT'95.

\bibitem{ChHa80}
A.~K. Chandra and D.~Harel.
\newblock {Computable Queries for Relational Databases}.
\newblock {\em Journal of Computer and System Sciences}, 21:156--178, 1980.

\bibitem{GrGrZu01}
G.~Greco, S.~Greco, and E.~Zumpano.
\newblock {A Logic Programming Approach to the Integration, Repairing and
  Querying of Inconsistent Databases}.
\newblock In {\em International Conference on Logic Programming}, pages
  348--364. Springer-Verlag, LNCS 2237, 2001.

\bibitem{GrZu00}
S.~Greco and E.~Zumpano.
\newblock {Querying Inconsistent Databases}.
\newblock In {\em International Conference on Logic for Programming and
  Automated Reasoning}, pages 308--325. Springer-Verlag, LNCS 1955, 2000.

\bibitem{Hull97}
R.~Hull.
\newblock {Managing Semantic Heterogeneity in Databases: A Theoretical
  Perspective}.
\newblock In {\em ACM Symposium on Principles of Database Systems}, pages
  51--61, 1997.
\newblock Invited talk.

\bibitem{Mot99}
A.~Motro.
\newblock {Multiplex: A Formal Model for Multidatabases and Its
  Implementation}.
\newblock In {\em International Workshop on Next Generation Information
  Technology and Systems}, pages 138--158. Springer-Verlag, LNCS 1649, 1999.

\bibitem{Var82}
M.~Y. Vardi.
\newblock {The Complexity of Relational Query Languages}.
\newblock In {\em ACM Symposium on Theory of Computing}, pages 137--146, 1982.

\end{thebibliography}
\end{document}